\documentclass[twocolumn]{aastex63}
\usepackage{amsmath}

\newcommand{\msun}{${\rm M_{\sun}}$}
\def\ltsima{$\; \buildrel < \over \sim \;$}
\def\simlt{\lower.5ex\hbox{\ltsima}}
\def\gtsima{$\; \buildrel > \over \sim \;$}
\def\simgt{\lower.5ex\hbox{\gtsima}}
\def\km{{\rm\,km}}
\def\kms{{\rm\,km\,s^{-1}}}
\def\km2s2{{\rm\,km^2\,s^{-2}}}

\def\kpc{{\rm\,kpc}}

\def\msun{{\rm\,M_\odot}}


\def\Gyr{{\rm\,Gyr}}

\def\ltsima{$\; \buildrel < \over \sim \;$}
\def\gtsima{$\; \buildrel > \over \sim \;$}

\def\Gaia{{\it Gaia\,\,}}
\def\Pontus{{\it Pontus}}
\def\GSE{{\it Gaia-Sausage/Enceladus}}

\submitjournal{ApJL}

\shorttitle{Characterization of Pontus}
\shortauthors{Malhan}

\begin{document}

\title{A new member of the Milky Way's family tree:\\ Characterizing the Pontus merger of our Galaxy}

\correspondingauthor{Khyati Malhan}
\email{kmalhan07@gmail.com}

\author[0000-0002-8318-433X]{Khyati Malhan}
\affiliation{Humboldt Fellow and IAU Gruber Fellow}
\affiliation{Max-Planck-Institut f\"ur Astronomie, K\"onigstuhl 17, D-69117, Heidelberg, Germany}

\begin{abstract}
We study the $\textit{Pontus}$ structure -- a recently discovered merger that brought in $\sim7$ globular clusters in the course of the hierarchical build-up of the Milky Way's halo. Here, we analyse the stellar population of $\textit{Pontus}$ and examine (1) its phase-space distribution using the ESA/$\textit{Gaia}$ dataset, (2) its metallicity and chemical abundances (i.e., [Fe/H], [$\alpha$/Fe], [Mg/Fe], [Al/Fe]) using the spectroscopic catalogue of APOGEE~DR17, and (3) the color-magnitude diagram that shows interesting features, including a possibly double horizontal branch and a small population of blue stragglers. In sum, the $\textit{Pontus}$ stars show some unique properties that suggest they likely originated from the merging of an independent satellite galaxy; future analysis will shed more light on the true nature of this structure. This chemo-dynamical analysis of $\textit{Pontus}$ stars is another step forward in our bigger quest to characterize $\textit{all}$ the merging events of our Milky Way.
\end{abstract}

\section{Introduction}\label{sec:Introduction}

The advent of the data of the \Gaia\ mission \citep{2016A&A...595A...1G, 2018A&A...616A...2L, 2021A&A...649A...1G} has allowed us to detect and characterize many sub-structures in the Milky Way halo \citep{Belokurov2018, Helmi2018, Myeong2019, Matsuno2019, Koppelman2019, Mackereth_2019, Yuan2020a, Ibata_2020_Sgr, Ramos_2020_Sgr, Naidu2020, Malhan_2021_LMS1}. These sub-structures are remnants of those progenitor galaxies that merged with the Milky Way and contributed to the stellar (and dark matter) population of the Galactic halo. Here, our aim is to specifically analyse one of these sub-structures, namely \Pontus\footnote{In Greek mythology, ``\Pontus'' (meaning ``the Sea'') is the name of one of the first children of the Gaia deity.}, and study the dynamical and chemical properties of its stellar population.

\begin{figure*}
\begin{center}
\includegraphics[width=\hsize]{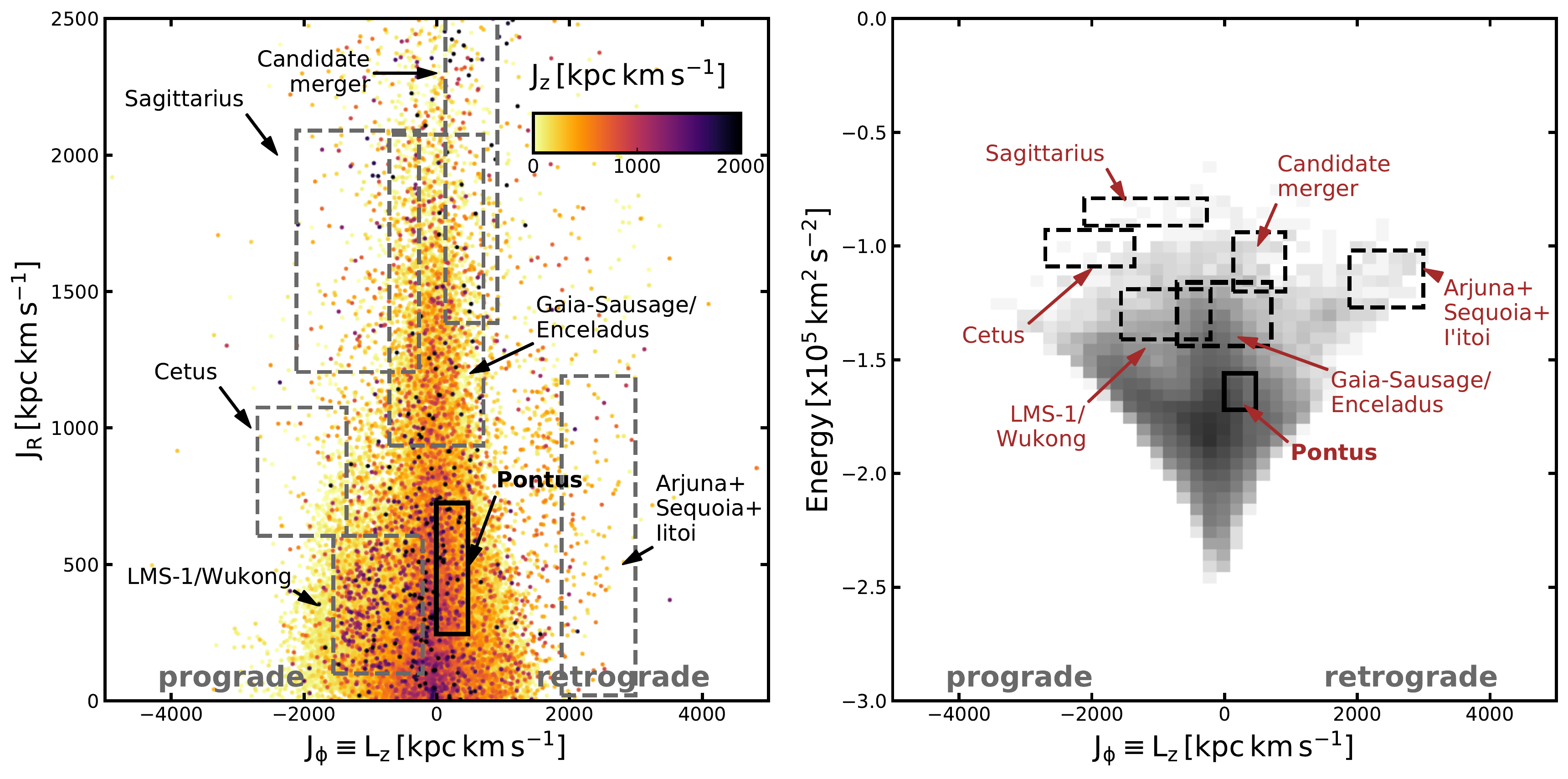}
\end{center}
\vspace{-0.4cm}
\caption{Action-Energy ($\mathbf{J}, E$) space of the Milky Way showing stars that form part of our halo sample (see Section~\ref{sec:Data} and \ref{sec:Orbits}). The left panel shows $J_\phi (\equiv L_z)$ vs $J_R$, where stars are colored by their $J_z$ values. The right panel shows $J_\phi$ vs $E$, that is represented by a 2D histogram colored by the logarithm of the number of stars per bin. In both the panels, the boxes describe the dynamical volumes occupied by different mergers as per \cite{Malhan_2022}. Stars belonging to \Pontus\ possess low energy ($E\sim-1.65\times10^5 \km2s2$) and slightly retrograde motion ($J_\phi \sim200 \kpc\kms$).}
\label{fig:Fig_1}
\end{figure*}

\Pontus\ was recently discovered by \cite{Malhan_2022} as a grouping of $7$~globular clusters that are tightly-clumped in the action-energy ($\mathbf{J},E$) space of the Milky Way. This group possesses low energy (i.e., the orbits of its components are more tightly bound to the Milky Way compared to other merger groups) and slightly retrograde motion (this is also shown in Figure~\ref{fig:Fig_1} that we explain below). The member globular clusters include NGC~288, NGC~5286, NGC~7099/M~30, NGC~6205/M~13, NGC~6341/M~92, NGC~6779/M~56, NGC~362; there exist two additional (tentative) associations, namely NGC~6864/M~75 and NGC~7089/M~2. In addition to being tightly-clumped in the ($\mathbf{J},E$) space, these globular clusters also show a tight age-metallicity relation -- again supporting the accretion scenario from a gradually enriched satellite galaxy.  

While \cite{Malhan_2022} could only analyse the globular cluster population of \Pontus, our central objective here is to analyse its general, unbound stellar population. This is important to understand, for instance: how are the \Pontus\ stars distributed in the phase-space of the Milky Way? What are their metallicity ([Fe/H]) and chemical abundances ([X/Fe])? What can we learn about this sub-structure by inspecting its color-magnitude diagram (CMD)? When did the (hypothesized) \Pontus\ galaxy merge with the Milky Way? Here, we make a first attempt to answer these questions by analysing the \Pontus\ stars.

This article is structured as follows. In Section~\ref{sec:Data}, we first describe our procedure to construct a sample of halo stars using the \Gaia\ dataset. Next, in Section~\ref{sec:Orbits}, we compute the orbits of these stars to specifically select the \Pontus\ population. In Section~\ref{sec:Analysis}, we analyse the \Pontus\ population to examine its phase-space distribution, chemical abundances and CMD. Finally, we discuss our findings and conclude in Section~\ref{sec:Conclusion}.

\section{Selecting the halo stars using Gaia}\label{sec:Data}

We first construct a sample of ``halo'' stars  -- those stars that possess kinematic very different from the local disk stars. For this, we use the \Gaia\ EDR3 dataset, but consider only those stars that possess radial velocity measurements from the \Gaia\ DR2 dataset. This dataset provides 6D phase-space information for $7,209,831$~sources in the heliocentric coordinates: on-sky positions ($\alpha,\delta$), parallaxes ($\varpi$), proper motions ($\mu^{*}_{\alpha},\mu_\delta$) and line-of-sight velocities ($v_{\rm los}$), along with their associated uncertainties. This dataset also provides the $G$, $G_{\rm BP}$ and $G_{\rm RP}$ magnitudes. 

To select the halo population, we follow a similar prescription described by \cite{Koppelman_2019_Helmi}, but with slight modification. We select those stars that follow the criteria: (1) $|\mathbf{V}-\mathbf{V_{\rm LSR}}|\geq210\kms$; where $\mathbf{V}$ is the velocity vector of a given star in the Galactocentric cartesian coordinates (this we obtain by transforming the heliocentric coordinates into cartesian coordinates using the Sun's velocity from \citealt{Drimmel_2018} as $(V_{X,\odot}, V_{Y,\odot},V_{Z,\odot})=(12.9,245.6,7.78)\kms$ and the Sun's location from \citealt{Gravity2018} as $R_{\odot}=8.112\kpc$) and $\mathbf{V_{\rm LSR}}$ is the velocity vector of the local standard of rest (LSR) that we adopt from \cite{Drimmel_2018} as $(V_{\rm X,LSR}, V_{\rm Y,LSR}, V_{\rm Z, LSR})=(1.8, 233.4,0.53)\kms$. (2) $\varpi>0$; this condition ensures that the parallaxes can be directly inverted to obtain heliocentric distances as $D_\odot=1/\varpi$ (this condition was implemented after correcting for the parallax zero-point of each star using the code\footnote{See \url{https://gitlab.com/icc-ub/public/gaiadr3_zeropoint} .} provided by \citealt{Lindegren_2021}, and henceforth all the parallaxes are zero-point corrected). (3) \texttt{parallax\_over\_error}$\geq5$; this quality cut ensures that our resulting sample contains well measured parallaxes with relative parallax error of $\leq20\%$. (4) \texttt{phot\_bp\_rp\_excess\_factor}$\leq1.27$; this condition removes some of the not so well-behaved globular cluster stars, and hence, we do not apply a colour-dependent correction \citep{Arenou_2018}. (5) RUWE$<1.4$; this value has been prescribed as a quality cut for ``good'' astrometric solutions by \cite{Lindegren_2021}. We note that the RUWE parameter was not considered by \cite{Koppelman_2019_Helmi}, and has also been ignored by many previous studies that had the similar objective of constructing a halo sample.

This halo selection renders $41,861$ data points. In other words, $\sim0.5\%$ of our \Gaia\ sample corresponds to the halo population; at least as per our definition of halo.

%
\section{Computing the orbits}\label{sec:Orbits}

For the above halo stars, we compute their $\mathbf{J}, E$ and also other orbital parameters (e.g., apocentre, $r_{\rm apo}$, pericentre, $r_{\rm peri}$, eccentricity, $ecc$). This is a necessary first step to particularly select the \Pontus\ stars using a specified range of orbital parameters; this range is described below. To compute the orbits, we adopt the Galactic potential model of \cite{McMillan2017}; similar to that used by \cite{Malhan_2022}. This is a static and axisymmetric model comprising a bulge, disk components and an NFW halo. To set this potential model, and to compute the orbital parameters, we make use of the \texttt{galpy} module \citep{Bovy2015}. 

Figure~\ref{fig:Fig_1} shows the resulting $\mathbf{J}$ distribution (left panel) and $(J_\phi,E)$ distribution (right panel) of all the stars from our halo sample. The actions are represented in the cylindrical coordinates, i.e., $\mathbf{J}\equiv(J_R,J_\phi,J_z)$, where $J_\phi$ corresponds to the z-component of angular momentum (i.e., $J_\phi\equiv L_z$) and negative $J_\phi$ represents prograde motion. Components $J_R$ and $J_z$ describe the extent of oscillations in cylindrical radius and $z$ directions, respectively. In this figure, the boxes are drawn using the prescription provided by \cite{Malhan_2022} and they describe the dynamical volumes occupied by different merger groups. 

We particularly focus on \Pontus\ that corresponds to the dynamical region described by: $E=[ -1.72 , -1.56 ]\times10^5\km2s2$, $J_R=[ 245 , 725 ]\kms\kpc$, $J_\phi= [ -5 , 470 ]\kms\kpc$, $J_z= [ 115 , 545 ]\kms\kpc$, $ecc= [ 0.5 , 0.8 ]$, $ r_{\rm peri} = [ 1 , 3 ]\kpc$, $ r_{\rm apo} = [ 8 , 13 ]\kpc$. These dynamical ranges are described using the values provided by \cite{Malhan_2022}, who found that the member globular clusters of \Pontus\ lie in this range\footnote{Since the member globular clusters of \Pontus\ are contained within this dynamical range, it is reasonable to assume that a significant fraction of its stellar population must also be contained within this range. Although, we note that the ``true'' extent out to which the \Pontus\ stars are distributed in the phase-space of the Milky Way could be larger.}. This implies that \Pontus\ stars are located in the  low-energy part of the halo, and they possess slightly retrograde motion and moderately eccentric orbits. We employ these strict dynamical parameters to make a box selection that yields $1311$ stars, and we refer to this sample as \Pontus\footnote{The reason for making cuts in this $7$ (non-independent) dimensions is to obtain a high-confidence (even though restricted) sample of \Pontus\ stars. While one may argue that making cuts only in the $\mathbf{J}$ space should be sufficient (since an orbit only has three integrals of motions), we supplement this criteria with additional orbital parameters in order to minimise the contamination. Although, even this restricted selection results in a few contaminants in our \Pontus\ sample; as we see below.}. 

\begin{figure*}
\begin{center}
\includegraphics[width=\hsize]{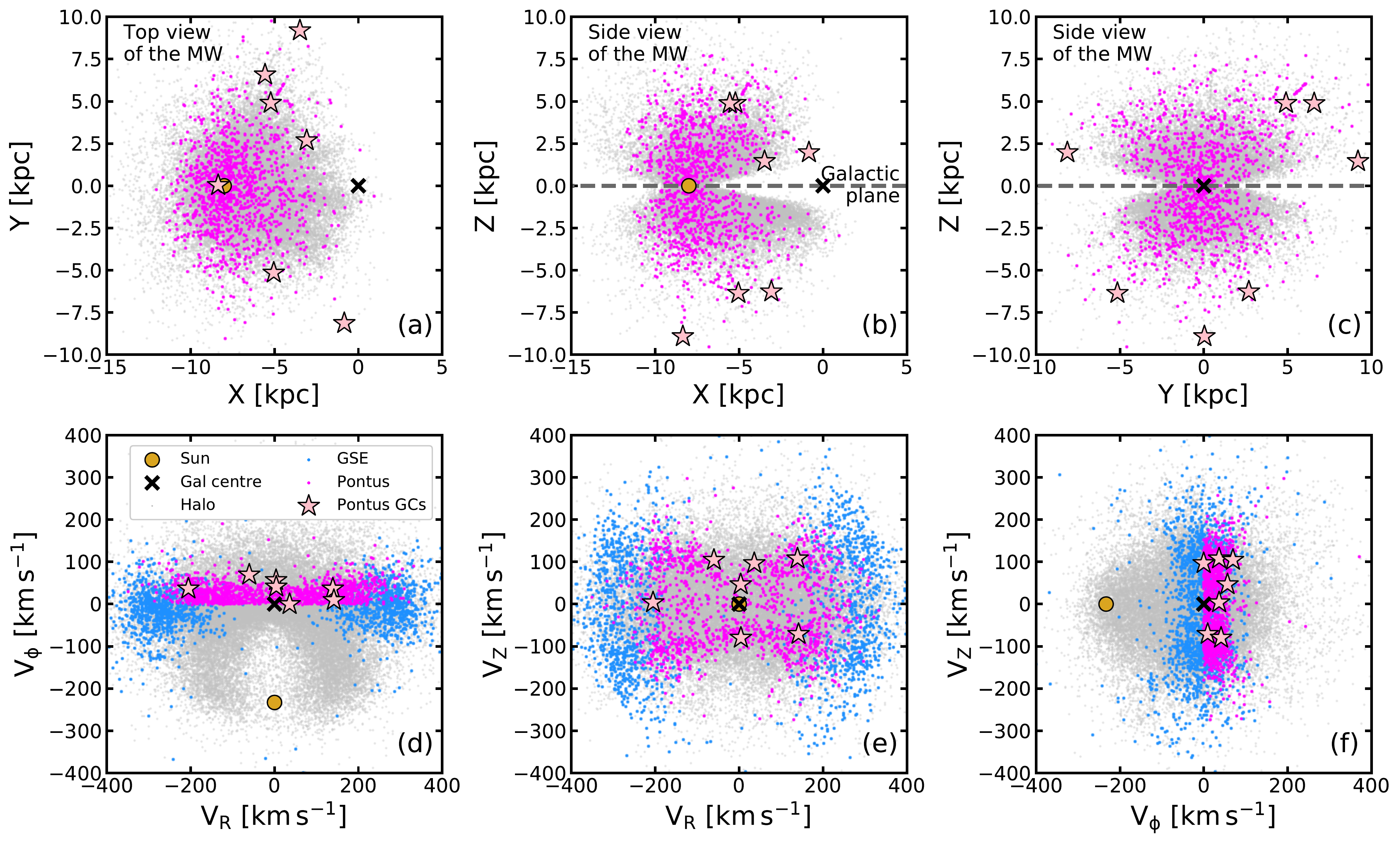}
\end{center}
\vspace{-0.4cm}
\caption{Phase-space distribution of the \Pontus\ stars (in ``magenta'' color) in the Galactocentric coordinates. The top panels show the spatial distribution of stars and the bottom panels show the velocity distribution. The ``star'' markers correspond to the member globular clusters of \Pontus. For comparison, we show the halo sample (in ``gray'') and also the \GSE\ stars (in ``blue'').}
\label{fig:Fig_2}
\end{figure*}
\section{Analysing the Pontus stars}\label{sec:Analysis}

We now analyse the \Pontus\ stars to examine their phase-space distribution in the Milky Way (Section~\ref{subsec:phase_space}), their metallicity and chemical abundances (Section~\ref{subsec:FeH}) and their CMD (Section~\ref{subsec:stellar_pop}).

\subsection{Phase-space distribution}\label{subsec:phase_space}

Figure~\ref{fig:Fig_2} shows the phase-space distribution of the \Pontus\ stars. 

In Figure~\ref{fig:Fig_2}, panels $a-c$ show the spatial distribution of the \Pontus\ stars in the Galactocentric cartesian coordinates. For comparison, we also show the stars from our halo sample. First, we note that \Pontus\ shows far fewer stars at small Galactocentric radii, although there is no obvious signature of a spatial overdensity. This implies that the \Pontus\ stars are almost completely phase-mixed in the Galaxy. This argument is further supported by the fact that these stars possess small $E$ values, implying that they orbit the inner region of the Galaxy which has small mixing timescales. In this figure, we also show the member globular clusters of \Pontus\ (as found by \citealt{Malhan_2022}). At first, it may seem strange that the clusters are distributed out to large heliocentric distances compared to the stars. However, this is because our \Gaia\ dataset contains very bright stars (with $G\simlt14$~mag\footnote{This is because radial velocity measurements are present for only those sources in \Gaia\ DR2 that are brighter than $G\approx14$~mag.}), and therefore, many of the distant (and fainter) stars are missing from our sample.

In Figure~\ref{fig:Fig_2}, panels $d-f$ show the velocity distribution of stars in the Galactocentric cylindrical coordinates. First, it can be seen that \Pontus\ stars occupy very limited region in this space -- very similar to the distribution of simulated sub-structures in \cite{Helmi_2000}. The reason that the \Pontus\ stars appear so confined in the velocity space is admittedly because of our narrow \Pontus\ (vs. halo) selection in the orbital parameter space. Secondly, the \Pontus\ population is stretched along the $V_R$ direction (and also along the spherical $V_r$ direction, that we independently checked), implying that these stars move along radial orbits; as expected from their moderately high $J_R$ and $ecc$ values (Figure~\ref{fig:Fig_1}, left panel). Moreover, for the \Pontus\ stars, we estimate the average and standard deviation of their $V_\phi$ distribution as $28 \pm 25 \kms$; implying a retrograde structure with fairly low dispersion in $V_\phi$. Furthermore, we note that all the member globular clusters overlap with the stars in this velocity space, and this is expected because of the $\mathbf{J}$, $E$ cut that we made to select the \Pontus\ stars.

We also make comparison of the \Pontus\ stellar population with that of \GSE\ \citep{Belokurov2018, Helmi2018} because these two structures are located close to each other in the ($J_\phi, E$) space and we are interested to examine -- do these two structures represent different merging events, or they simply correspond to two different fragments of the same merger? In this regard, \cite{Malhan_2022} argues that \Pontus\ and \GSE\ are independent structures, since the member globular clusters of each group occupy different locations in the ($\textbf{J}, E$) space (and consequently in phase-space) and they also possess different age-metallicty relation. Here, we want to examine whether this scenario is also favoured by their stellar population.

The sample of \GSE\ population is constructed in the same way as described for \Pontus\ in Section~\ref{sec:Orbits}, except using a different set of dynamical range as described by \cite{Malhan_2022}: $E\sim[ -1.44 , -1.16 ]\times10^5\km2s2$, $J_R\sim [ 935 , 2075 ]\kms\kpc$, $J_\phi\sim [ -715 , 705 ]\kms\kpc$, $J_z\sim [ 85 , 1505 ]\kms\kpc$, $ecc\sim [ 0.7 , 0.9 ]$, $ r_{\rm peri}\sim [ 1 , 4 ]\kpc$, $ r_{\rm apo}\sim [ 16 , 30 ]\kpc$. We note that this $E$ and $J_\phi$ range, that has been independently proposed by \cite{Malhan_2022}, is similar to that previously used by \cite{Koppelman2019} to select and study the \GSE\ stars. These stars are shown in panels $d-f$ of Figure~\ref{fig:Fig_2}. As can be seen, the \GSE\ and \Pontus\ stars occupy very different regions in the velocity space, although with small overlapping in the $V_\phi-V_z$ space. Previous studies that made similar $\mathbf{J}$ and $E$ selections to simultaneously analyse multiple structures have found that different structures can in fact highly overlap in the velocity space (e.g., \citealt{Koppelman2019}). Furthermore, for \GSE\ we estimate $V_\phi=-8 \pm 55 \kms$, suggesting that this population has almost no retrgroade component and possesses larger $V_\phi$ dispersion than \Pontus. 

\begin{figure*}
\begin{center}
\vbox{
\includegraphics[width=\hsize]{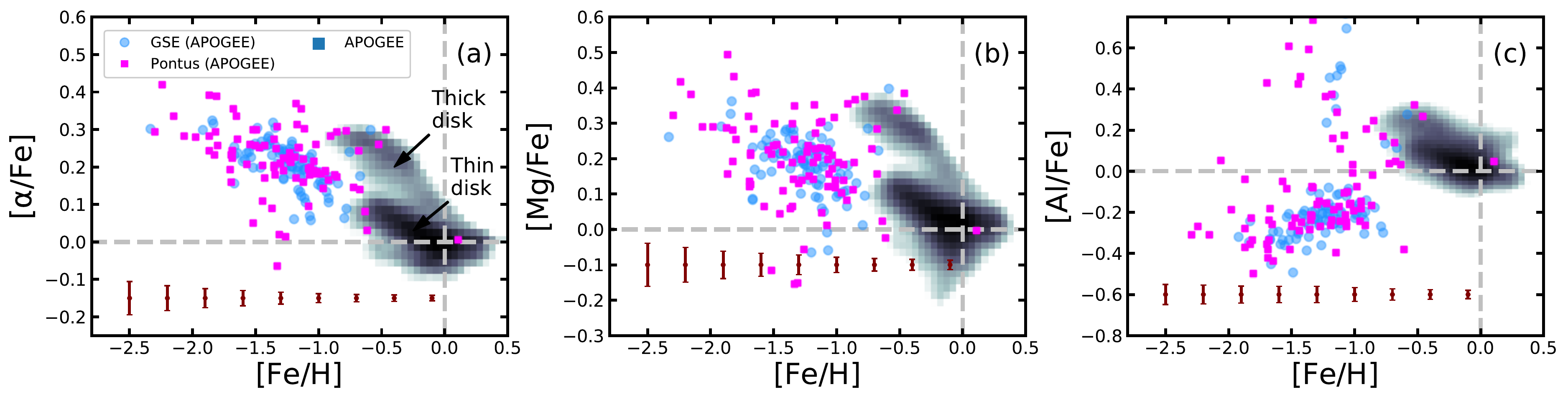}
}
\end{center}
\vspace{-0.5cm}
\caption{Analysing chemical abundances of \Pontus\ stars using the APOGEE dataset. Panel `a' shows the [Fe/H] vs. [$\alpha/$Fe] distribution. The background shows 2D histograms constructed using the entire APOGEE dataset (darker the bin, higher is the number of stars in that bin). The bars at the bottom of the panel indicate the median error at that [Fe/H]. For comparison, we also show the \GSE\ stars. Panel `b' shows the [Fe/H] vs. [Mg/Fe] distribution and panel `c' shows the [Fe/H] vs. [Al/Fe] distribution.}
\label{fig:Fig_3}
\end{figure*}
\subsection{Metallicity and chemical abundances}\label{subsec:FeH}

We now analyse the metallicity and chemical abundances for the \Pontus\ stars. To this end, we cross-match our sample with the APOGEE~DR17 catalogue \citep{Abdurrouf_2021} and find $88$ matches. We particularly analyse the following quantities: [Fe/H], [$\alpha$/Fe], [Mg/Fe] and [Al/Fe]. 

The \Pontus\ stars are shown in Figure~\ref{fig:Fig_3}. From the [Fe/H] vs. [$\alpha$/Fe] distribution, we note that the \Pontus\ stars clearly show an anti-correlation sequence that is indicative of their accretion (dwarf galaxy) origin. A few of the \Pontus\ stars overlap with the disk region, suggesting that our sample contains some disk contaminants. Secondly, we observe a similar anti-correlation sequence in [Fe/H] vs. [Mg/Fe], which is expected since Mg is an $\alpha$ element. On the other hand, a correlation sequence is observed in [Fe/H] vs. [Al/Fe]. This distribution reveals that some of the \Pontus\ stars overlap with the halo region, suggesting that our sample contains some halo contaminants.

For the \Pontus\ stars, we compute the means and dispersions of their distributions in [Fe/H], [$\alpha$/Fe], [Mg/Fe] and [Al/Fe], and these are provided in Table~\ref{tab:table_FeH}. These values are computed following our own Metropolis-Hastings based MCMC algorithm, based on \cite{Malhan_2021_LMS1}, where the log-likelihood function is taken to be:
\begin{equation}\label{eq:chemistry}
\ln \mathcal{L} = \sum_{\rm data} \Big[ -\ln \sigma_{\rm Y_{obs}} -0.5 \Big(\dfrac{\rm Y_0-Y_{\rm d}}{\sigma_{\rm Y_{obs}}}\Big)^2 \Big]\,.
\end{equation}
Here, $Y$ corresponds to the quantity of interest (e.g., [Fe/H] or [X/Fe]), ${\rm Y_{d}}$ is the measured quantity, and the Gaussian dispersion $\sigma_{\rm Y_{obs}}$ is the sum in quadrature of the intrinsic dispersion of the model together with the observational uncertainty of each data point ($\sigma^2_{\rm Y_{obs}}=\sigma^2_{\rm Y} + \sigma^2_{\rm Y_d}$). For this inference, we first impose conservative cuts so that our resulting sub-sample has a reduced contamination. In doing so, we retain only those stars that possess [Fe/H]$<-0.75$ (to reduce disk contaminants), [$\alpha$/Fe]$>0$ and [Al/Fe]$<0$ (to reduce halo contaminants). This selection renders $60$ stars, and this sub-sample is used to compute the values provided in Table~\ref{tab:table_FeH}. Table~\ref{tab:table_FeH} quotes the mean and the standard deviation corresponding to the posterior PDFs of different quantities.

\begin{table}
\centering
\caption{Metallicity and chemical abundances of the \Pontus, \GSE\ and halo populations.}
\label{tab:table_FeH}
\begin{tabular}{|l|c|c|c|}

\hline
\hline
Param.     & {\bf Pontus} & {\bf GSE} & {\bf Halo}\\ 
\hline
\hline

$\rm{[Fe/H]}$  & $ -1.403 \pm 0.047 $ & $ -1.274 \pm 0.03 $ & $ -1.185 \pm 0.009 $\\
$\rm{\sigma_{[Fe/H]}}$  & ($ 0.368 \pm 0.035 $) & ($ 0.253 \pm 0.021 $) & ($ 0.497 \pm 0.007 $)\\

& & & \\            
$\rm{[\alpha/Fe]}$ & $ 0.235 \pm 0.008 $ & $ 0.216 \pm 0.006 $ & $ 0.239 \pm 0.002 $\\
$\rm{\sigma_{[\alpha/Fe]}}$  & ($ 0.057 \pm 0.006 $) & ($ 0.05 \pm 0.005 $) & ($ 0.085 \pm 0.001 $)\\

& & & \\                      
$\rm{[Mg/Fe]}$   & $ 0.211 \pm 0.012 $ &  $ 0.193 \pm 0.008 $ & $ 0.246 \pm 0.002 $\\
$\rm{\sigma_{[Mg/Fe]}}$    & ($ 0.09 \pm 0.01 $)
 & ($ 0.06 \pm 0.006 $) & ($ 0.111 \pm 0.002 $)\\

& & & \\            
$\rm{[Al/Fe]}$ & $ -0.228 \pm 0.014 $ & $ -0.237 \pm 0.009 $ & $ -0.037 \pm 0.005 $\\
$\rm{\sigma_{[Al/Fe]}}$ & ($ 0.103 \pm 0.011 $) & ($ 0.074 \pm 0.007 $) & ($ 0.285 \pm 0.004 $)\\

& & & \\
\hline
\hline
\end{tabular}
\tablecomments{From left to right the columns provide the parameter of interest, measurements corresponding to the \Pontus, \GSE\ and halo populations.}
\end{table}
\begin{table}
\centering
\caption{Results corresponding to the two-sample Kolmogorov-Smirnov (KS) test performed on \Pontus\ and halo populations, \GSE\ and halo populations, and \Pontus\ and \GSE\ populations.}
\label{tab:table_comp}
\begin{tabular}{|l|c|c|c|}

\hline
\hline
Param. & Pontus-Halo & GSE-Halo & Pontus-GSE\\
\hline
\hline

$\rm{[Fe/H]}$   &  $ <0.001 \,(> 4.5 \sigma)$  &  $ <0.001 \,(> 5.0 \sigma)$  &  $ 0.028 \,(> 2.0 \sigma)$  \\

$\rm{[\alpha/Fe]}$   &  $ <0.001 \,(> 3.5 \sigma)$  &  $ <0.001 \,(> 5.5 \sigma)$  &  $ 0.414 \,(> 0.5 \sigma)$  \\

$\rm{[Mg/Fe]}$   &  $ <0.001 \,(> 4.5 \sigma)$  &  $ <0.001 \,(> 7.0 \sigma)$  &  $ 0.254 \,(> 0.5 \sigma)$  \\

$\rm{[Al/Fe]}$   &  $ <0.001 \,(> 6.5 \sigma)$  &  $ <0.001 \,(> 8.0 \sigma)$  &  $ 0.358 \,(> 0.5 \sigma)$  \\

& & & \\
\hline
\hline
\end{tabular}
\tablecomments{Left most column provides the name of the parameter, the next $3$ columns give p-values of the KS test (and the corresponding significance at which the null hypothesis can be rejected).}
\end{table}

For comparison, we also show the \GSE\ stars in Figure~\ref{fig:Fig_3}. These stars appear to follow similar sequences as that of \Pontus\ in [Fe/H] vs. [$\alpha$/Fe] and [Fe/H] vs. [Mg/Fe], except they are slightly less scattered. The measured parameters for \GSE\ stars are provided in Table~\ref{tab:table_FeH}, and these were computed in a similar way as described above for the \Pontus\ population (where we imposed similar quality cuts to construct the sub-sample). Table~\ref{tab:table_FeH} also shows the measurements for our halo sample. These measurements were obtained by first cross-matching our halo sample with the APOGEE dataset, and then using equation~\ref{eq:chemistry}. In doing so, note that we did not impose any quality cuts on [Fe/H], [$\alpha$/Fe] or [Al/Fe] as above. This analysis indicates that the \Pontus\ population possesses somewhat similar distribution in [$\alpha$/Fe] as that of the halo population, while they differ in [Fe/H], [Mg/Fe] and [Al/Fe]. Furthermore, the comparison between the \Pontus\ and \GSE\ populations indicate that while their distributions are somewhat similar in [$\alpha$/Fe], [Mg/Fe] and [Al/Fe], they slightly differ in [Fe/H]. In particular, the \Pontus\ stars on an average are slightly more metal-poor than the \GSE\ stars. 

Table~\ref{tab:table_comp} summarises the result from the Kolmogorov-Smirnov (KS) test that we perform to compare the [Fe/H], [$\alpha$/Fe], [Mg/Fe] and [Al/Fe] distributions of \Pontus\ and \GSE, \Pontus\ and halo, and \GSE\ and halo. The KS test is performed for the null hypothesis that the two given samples are drawn from the same distribution. For instance, a value of $>5\sigma$ in Table~\ref{tab:table_comp} implies that the null hypothesis can be rejected at the $>5\sigma$ level\footnote{Similar KS-test has also been previously used to examine chemical distributions of different halo sub-structures (e.g., \citealt{Malhan_2019_Kshir}).}. This comparison shows: (1) The [Fe/H], [$\alpha$/Fe], [Mg/Fe] and [Al/Fe] distributions of both \Pontus\ and \GSE\ is very different than that of the halo population. (2) For the \Pontus\ and \GSE\ populations, their distributions in [$\alpha$/Fe], [Mg/Fe] and [Al/Fe] are quite similar (as the null hypothesis can only be rejected at the $>0.5\sigma$ level). (3) The null hypothesis that -- the two [Fe/H] samples for \Pontus\ and \GSE\ are drawn from the same distribution -- can be rejected at the $>2\sigma$ level.

\begin{figure*}
\begin{center}
\includegraphics[width=\hsize]{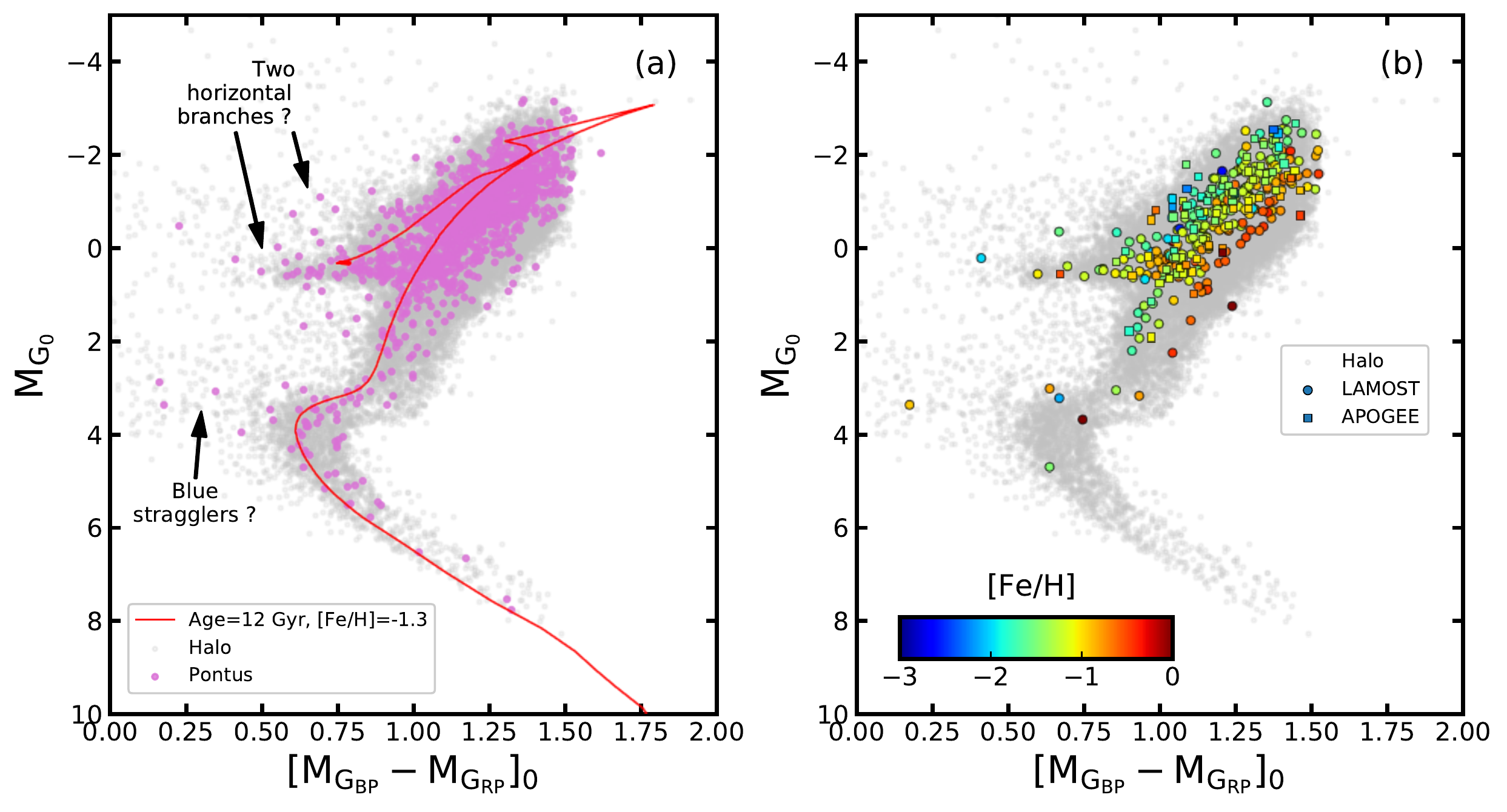}
\end{center}
\vspace{-0.5cm}
\caption{The CMD of the \Pontus\ stars. Panel `a' shows \Pontus\ stars in magenta and the halo population in gray. We also plot a stellar population model that is similar to the average metallicity of \Pontus. Panel `b' is similar to panel `a', except \Pontus\ stars are now colored by their [Fe/H] values. To make this plot, we also cross-match our \Pontus\ sample with the LAMOST~DR5 dataset \citep{Zhao_2012_LAMOST, Xiang_2019}.}
\label{fig:Fig_4}
\end{figure*}
\subsection{Stellar population}\label{subsec:stellar_pop}

Figure~\ref{fig:Fig_4}a shows the CMD of the \Pontus\ population. To construct this figure, we first correct \Gaia\ magnitudes for dust extinction using \cite{SFD_1998} maps and assuming the extinction ratios $A_G/A_V = 0.86117$, $A_{\rm BP}/A_V =1.06126$, $A_{\rm RP}/A_V = 0.64753$ (as listed on the web interface of the PARSEC isochrones, \citealt{Bressan_2012}). Henceforth, all magnitudes and colors refer to these extinction-corrected values. Next, we transform these corrected apparent magnitudes ($G_{\rm BP, 0}, G_{\rm RP,0}, G_0$) into absolute magnitudes ($M_{G_{\rm BP, 0}}, M_{G_{\rm RP, 0}}, M_{G_0}$) using heliocentric distances of stars ($D_\odot$). For comparison, we also show the CMD of the halo population. 

From Figure~\ref{fig:Fig_4}a, we first notice that the \Pontus\ population contains more of intrinsically brighter stars than fainter stars, and this is likely due to our quality cuts that can give rise to such asymmetries (e.g., \citealt{Koppelman_2019_Helmi}). The broadness of the \Pontus\ CMD indicates that these stars do not resemble a single stellar population, but rather favour a wide [Fe/H] distribution (as we also found in Section~\ref{subsec:FeH}) and possibly also wide age distribution. The mean of this age distribution is likely $\sim12\Gyr$ -- this point is illustrated by plotting an isochrone model corresponding to ([Fe/H], Age)=($-1.3$~dex, $12\Gyr$) that fits the CMD reasonably well. 

The CMD of \Pontus\ shows some interesting features. For instance, Figure~\ref{fig:Fig_4}a shows the possibility of two horizontal branches and a small population of blue stragglers. In order to examine whether these features are really representative of the \Pontus\ population, or they are the Milky Way interlopers, we inspect Figure~\ref{fig:Fig_4}b that shows the CMD colored by the [Fe/H] measurements. As can be seen, we do not possess substantial number of [Fe/H] measurements for the horizontal branch stars, and therefore, we can not confidently confirm the presence of a double horizontal branch. However, a few of these stars do possess [Fe/H] measurements that are consistent with the average [Fe/H] of \Pontus. On the other hand, for the blue stragglers, we possess [Fe/H] measurement for only one star, and it is slightly more metal-rich than the average [Fe/H] of \Pontus.

\section{Conclusion and Discussion}\label{sec:Conclusion}

We have performed the chemo-dynamical analysis of the stellar population of the recently discovered \Pontus\ structure. For this, we used the \Gaia\ dataset to analyse its phase-space distribution, and further used the APOGEE dataset to examine its metallicity and chemical abundances.

In regard to the dynamical properties: the \Pontus\ stars possess low-energy and slightly retrograde motion (see Figure~\ref{fig:Fig_1}b) and they move on moderately eccentric orbits. This implies that the (hypothesized) progenitor galaxy would have merged on a radial, and slightly retrograde, orbit. This argument is also supported by the fact that the \Pontus\ stars are stretched in the velocity space along the radial direction (see Figure~\ref{fig:Fig_2}). 

In regard to the metallicity and chemical abundances: the \Pontus\ population is metal-poor and $\alpha$-enhanced (see Table~\ref{tab:table_FeH}), that is indicative of its accretion (dwarf galaxy) origin. Based on this, we argue that \Pontus\ likely merged with the Milky Way $>5-6\Gyr$ ago. This can be argued using the observations of the {\it Sagittarius} merger \citep{Ibata1994}. For {\it Sagittarius}, we know that its stars possess [Mg/Fe]$=-0.03$ \citep{Hayes_2020_Sgr} and that this merging occured $\sim 5-6\Gyr$ ago \citep{Ruiz-Lara2020}. Given that \Pontus\ is more Mg-enhanced than {\it Sagittarius}, this tentatively suggests that the \Pontus\ galaxy must have accreted earlier than {\it Sagittarius} and therefore it did not get sufficient time to get enriched in Fe (via Type Ia supernovae) and ended up retaining more of Mg (and also other $\alpha$ elements). This early accretion scenario for \Pontus\ is also supported by its very low-energy, implying that the progenitor galaxy was likely accreted very early on when the Milky Way was much smaller in size (than today), and therefore, any merging galaxy would have ended up populating the very inner (low-energy) regions of the Galaxy. Moreover, given the [Fe/H] of \Pontus, it appears that the progenitor galaxy must have been a massive system (with $M_\star \sim 10^7\msun$; this value is obtained using \citealt{Kirby_2013} relation).

The CMD of the \Pontus\ stars shows some interesting features (Figure~\ref{fig:Fig_4}). These include a possibly double horizontal branch and a small popuation of blue stragglers. We note that similar twin horizontal branch feature has also been observed in the globular cluster NGC~6205/M~13 \citep{Grundahl_1998} that is associated with the \Pontus\ galaxy \citep{Malhan_2022}. Although, we note that \cite{Grundahl_1998} discusses $\sim0.4$~mag ``overluminous'' horizontal branch for only blue stars in M~13, and here we observe a gap of $\sim1$~mag between the two branches. In \Pontus, if the brighter branch is a real feature, it may correspond to the post horizontal branch stars (e.g., \citealt{Sandquist_2010, Davis_2022}). 

During our analysis, we also made comparisons between the stellar populations of \Pontus\ and \GSE. This was done because these two groups are located close to each other in the ($J_\phi, E$) space and we were interested to examine whether they represent different merging events or two fragments of the same merger. 

It is interesting to note both the differences and similarities between \Pontus\ and \GSE\ structures. First, Figure~\ref{fig:Fig_1} shows that -- while these two structures are located close to each other in the orbit-space, their difference in this dynamical space is still larger compared to the difference between other structures that are also located close to each other (e.g., {\it Sagittarius}--{\it Cetus} or {\it Cetus}--{\it LMS-1/Wukong} or {\it Arjuna}-{\it Sequoia}-{\it I'itoi}). Nonetheless, it is not unexpected for the stellar debris of a single massive merger to span a large range of distribution in $L_z-E$ space (e.g., \citealt{Helmi2018}). However, this last point is difficult to confirm because the true dynamical boundaries for \GSE\ and \Pontus\ are currently unknown (in terms of observations). Secondly, for \GSE, we deem that our orbit selection criteria (that is based on \citealt{Malhan_2022}) does not necessarily underestimate the dynamical extent of this structure, as previous studies have also used similar range of $L_z-E$ to select and study the \GSE\ population (e.g., \citealt{Koppelman2019}). Moreover, we find that these two populations show different behaviour in the velocity space (Figure~\ref{fig:Fig_2}); although, this is a consequence of our orbit selection criteria. Furthermore, while these two populations possess similar [$\alpha$/Fe], [Mg/Fe] and [Al/Fe] distributions, they somewhat differ in their [Fe/H] distributions (see Table~\ref{tab:table_FeH} and ~\ref{tab:table_comp}). Also, as shown in \cite{Malhan_2022}, the member globular clusters of \Pontus\ and \GSE\ possess slightly different age-metallicity relations.

For future analysis of \Pontus, it will be useful to run detailed N-body simulations to better understand how this merging event would have taken place, and to predict the present-day phase-space distribution of its stars in the Milky Way. Moreover, building chemical evolution model for \Pontus\ will be crucial to constrain its star formation history and the ages of its stars (e.g., \citealt{Vincenzo_2019}). These constraints, together, will also be useful to confirm more confidently whether \Pontus\ is the low-energy (low $J_R$) fragment of \GSE, or that they truly are the remnants of independent merging events. Furthermore, this will also shed light on the similar or distinct origin of \Pontus\ and {\it Thamnos} \citep{Koppelman2019}, that is another structure that lies close to \Pontus\ in the ($J_\phi,E$) space. These future steps should be easy to explore with the upcoming \Gaia\ DR3 RVS dataset (that will be $2$~mag deeper than DR2 RVS) and the spectroscopic datasets of WEAVE \citep{Dalton_2014} and 4MOST \citep{deJong_2019} surveys that will enable us to gather high quality phase-space and chemical abundance information for a larger sample of \Pontus\ stars. Therefore, future analysis of \Pontus, and its comparison with the other mergers, will allow us to gain a more comprehensive understanding of the formation of the Milky Way's vast stellar halo.

\section*{Acknowledgements}

We thank our referee for reviewing this manuscript. It is a pleasure to thank Hans-Walter Rix, Morgan Fouesneau, Jianhui Lian and Anke Arentsen for their help and suggestions. KM acknowledges support from the Alexander von Humboldt Foundation at Max-Planck-Institut f\"ur Astronomie, Heidelberg and is also grateful to the IAU's Gruber Foundation Fellowship Programme for their finanacial support. 

This work has made use of data from the European Space Agency (ESA) mission {\it Gaia} (\url{https://www.cosmos.esa.int/gaia}), processed by the {\it Gaia} Data Processing and Analysis Consortium (DPAC, \url{https://www.cosmos.esa.int/web/gaia/dpac/consortium}). Funding for the DPAC has been provided by national institutions, in particular the institutions participating in the {\it Gaia} Multilateral Agreement.


Funding for SDSS-III has been provided by the Alfred P. Sloan Foundation, the Participating Institutions, the National Science Foundation, and the U.S. Department of Energy Office of Science. The SDSS-III web site is http://www.sdss3.org/.

SDSS-III is managed by the Astrophysical Research Consortium for the Participating Institutions of the SDSS-III Collaboration including the University of Arizona, the Brazilian Participation Group, Brookhaven National Laboratory, Carnegie Mellon University, University of Florida, the French Participation Group, the German Participation Group, Harvard University, the Instituto de Astrofisica de Canarias, the Michigan State/Notre Dame/JINA Participation Group, Johns Hopkins University, Lawrence Berkeley National Laboratory, Max Planck Institute for Astrophysics, Max Planck Institute for Extraterrestrial Physics, New Mexico State University, New York University, Ohio State University, Pennsylvania State University, University of Portsmouth, Princeton University, the Spanish Participation Group, University of Tokyo, University of Utah, Vanderbilt University, University of Virginia, University of Washington, and Yale University.

\bibliography{ref1}
\bibliographystyle{aasjournal}

\end{document}